\documentclass[letterpaper]{article}

\usepackage{palatino}

\usepackage[USenglish]{babel}

\usepackage{amsmath,amsthm,amsfonts,amssymb}
\usepackage{xfrac}

\usepackage{geometry}
\usepackage{hyperref}

\usepackage{graphicx}
\usepackage{color}

\usepackage{setspace}
\begin{document}

\begin{titlepage}
	
\vspace*{\stretch{1}}
	
\begin{center}
	
\Large\textbf{Global Sensitivity Analysis in a Mathematical Model of the Renal Interstitium} \\
   
\vspace*{\stretch{1}}
   
\large\textit{
		Mariel Bedell, Carnegie Mellon University\\
		Claire Yilin Lin, Emory University\\
		Emmie Rom\'an-Mel\'endez, University of Puerto Rico Mayaguez\\
		Ioannis Sgouralis, National Institute for Mathematical and Biological Synthesis}      
\end{center}

\vspace*{\stretch{3.0}}

\begin{abstract}

The pressure in the renal interstitium is an important factor for normal kidney function. Here we develop a computational model of the rat kidney and use it to investigate the relationship between arterial blood pressure and interstitial fluid pressure. In addition, we investigate how tissue flexibility influences this relationship. Due to the complexity of the model, the large number of parameters, and the inherent uncertainty of the experimental data, we utilize Monte Carlo sampling to study the model's behavior under a wide range of parameter values and to compute first- and total-order sensitivity indices. Characteristically, at elevated arterial blood pressure, the model predicts cases with increased or reduced interstitial pressure. The transition between the two cases is controlled mostly by the compliance of the blood vessels located before the afferent arterioles.

\vspace{2ex}
\noindent{\bf Keywords:} mathematical model, sensitivity analysis, Monte Carlo, kidney, interstitium

\end{abstract}

\vspace*{\stretch{1.0}}

\end{titlepage}

%%%%%%%%%%%%%%%%%%%%%%%%%%%%%%%%%%%%%%%%%%%%%%%%%%%%%%%%%%%%%%%%
%%%%%%%%%%%%%%%%%%%%%%%%%%%%%%%%%%%%%%%%%%%%%%%%%%%%%%%%%%%%%%%%
\section{Introduction}

Kidneys are the core organs in the urinary system. Their principal functions are to remove meta\-bolic waste from the blood and to regulate blood salt and water levels \cite{Eaton2009Vanders-renal-p}. Through the regulation of salt and water, kidneys also play an important role in the regulation of arterial blood pressure \cite{Cowley1997Role-of-the-ren, Wolgast1981Functional-char}. To perform these functions, each kidney adjusts the composition of the urine it produces.

Each kidney has an outer layer, called the \emph{cortex}, and an inner layer, known as the \emph{medulla} \cite{Kriz1988A-standard-nome}. Much of the space in these regions is filled by the functional units of the kidney, which are termed \emph{nephrons}. Depending on the organism, each kidney contains thousands to millions of nephrons. Nephrons are responsible for the production of urine.

Kidneys contain two types of nephrons, cortical (short) and juxtamedullary (long) nephrons, each of which is surrounded by a net of capillaries. Cortical nephrons remain almost entirely in the cortex, while juxtamedullary nephrons extent deep into the medulla. Each nephron consists of a \emph{glomerulus} and a \emph{renal tubule}. Further, each renal tubule consists of various permeable or impermeable segments \cite{Eaton2009Vanders-renal-p,Kriz1988A-standard-nome}. Additionally, each nephron has access to a collecting duct for removal of the produced urine.

Kidneys are connected with the rest of the body by two blood vessels, the renal artery, which carries blood into the kidney, and the renal vein, which carries blood out of the kidney to recirculate the body. In addition, urine is excreted from the body through the ureter. Blood coming from the renal artery is delivered to the afferent arterioles. A steady flow of blood coming from the afferent arteriole of a nephron is filtered in the glomerulus and flows into the renal tubule. The blood flow is maintained constant in each glomerulus by the constriction or relaxation of its afferent arteriole \cite{Holstein-Rathlou1994Renal-blood-flo,Sgouralis2015Mathematical-mo}. Nearly all of the fluid that passes through the renal tubules is reabsorbed and only a minor fraction results in urine. Fluid is reabsorbed from the renal tubules in two stages: first by the renal interstitium and then by the surrounding capillaries. The processes underlying reabsorption are driven by the pressures in the interstitial spaces \cite{Cowley1997Role-of-the-ren, Wolgast1981Functional-char}.

Although the pressures in the renal interstitium are important determinants of kidney function, there is a lack of investigations that look at the factors affecting them. Here we develop a computational model of the rat kidney, for which several experimental data exist, and use it to study the relationship between arterial blood pressure and interstitial fluid pressure. In addition, we study how tissue flexibility affects this relationship and how the model predictions are affected by the uncertainty of key model parameters. We model the uncertain parameters as random variables and quantify their impact using Monte Carlo sampling and global sensitivity analysis.

%%%%%%%%%%%%%%%%%%%%%%%%%%%%%%%%%%%%%%%%%%%%%%%%%%%%%%%%%%%%%%%%
%%%%%%%%%%%%%%%%%%%%%%%%%%%%%%%%%%%%%%%%%%%%%%%%%%%%%%%%%%%%%%%%
\section{Methods}

\subsection{Model Description}

\begin{figure}[tb]
        \includegraphics[width=0.75\textwidth]{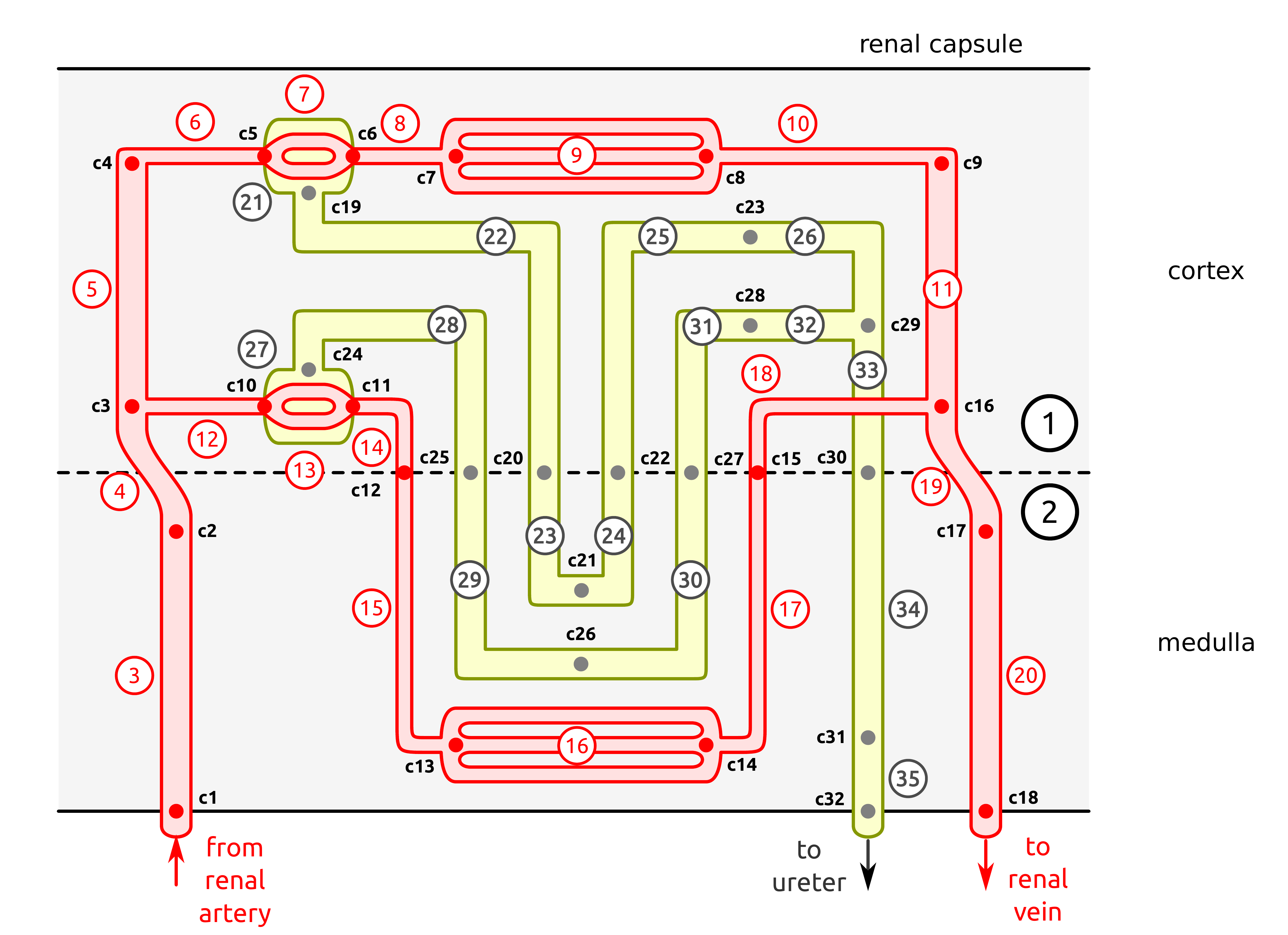}
        \centering
        \caption{Schematic diagram of the model kidney. The diagram shows the arrangement of blood vessels (red) and nephrons (yellow) within the interstitial spaces (grey). With the exceptions of the capillaries, the schematic displays only one of each of the different compartments contained in the full model. Nodes c1--c32 mark the connections of the compartments. For details see main text and Table~\ref{tb:1}.        }
        \label{NephronModel}
\end{figure}

The model consists of a collection of compartments that follow the characteristic anatomy of the kidneys of mammals \cite{Kriz1988A-standard-nome,Moffat2001A-vascular-patt}. The compartments fall in three categories: (i) \emph{regions} that model the cortical and medullary interstitial spaces, (ii) \emph{pipes} that model the blood vessels and renal tubules, and (iii) \emph{spheres} that model the glomeruli. A schematic diagram depicting the arrangement of the compartments (1--35) is shown on Figure~\ref{NephronModel} and a summary is given in Table~\ref{tb:1}. To facilitate the description of the model equations below, we use a set of nodes (c1--c32) that mark the connections of the compartments; these nodes are also included in Figure~\ref{NephronModel} and Table~\ref{tb:1}.

Briefly speaking, blood enters through the renal artery (node c1) and splits into a number of large arteries (compartments 3--5) that drain to the afferent arterioles (compartments 6 and 12). Each afferent arteriole supplies one glomerulus (compartments 21 and 27). In the glomeruli, blood is divided between the efferent arterioles (compartments 8 and 14) and the renal tubules (compartments 22--26 and 28--32). Leaving the efferent arterioles, blood passes through the cortical microcirculation (compartments 9 and 10) or the medullary microcirculation (compartments 15--18), before it rejoins in large veins (compartments 11, 19, 20) and leaves through the renal vein (node c18).

The model represents short (compartments 21--26) and long nephrons (compartments 27--32) that both drain in the same collecting duct (compartments 33--35), which, in turn, drains to the ureter (node c32). The model accounts for the spacial as well as the anatomical differences between the two nephrons that are developed in the mammalian kidney \cite{Kriz1988A-standard-nome,Moffat2001A-vascular-patt}. For example, the model accounts for differences in the location within the cortex or medulla, in the pre- and post-glomerular vascular supply, dimensions, reabsorptive capacity, etc.

\subsubsection{Model Pipes and Spheres}

\begin{figure}[tbp]
\centering
        \includegraphics [width=\textwidth ] {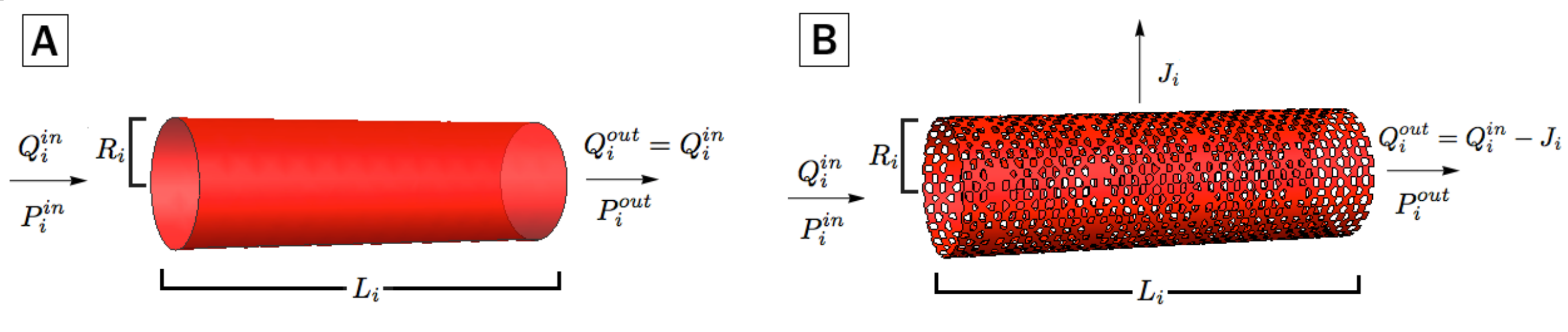}
        \caption{Model pipes. A: impermeable pipe. B: permeable pipe. For details see main text. }
        \label{pic:Cylinder}
\end{figure}

Blood vessels and renal tubules are modeled as distensible pipes. Glomeruli are modeled as distensible spheres. Fluid flows through a compartment $i$ at a volumetric rate of $Q_i$ (Figure~\ref{pic:Cylinder}). Following the physiology, some of the pipes are considered permeable while others impermeable \cite{Eaton2009Vanders-renal-p}. For simplicity, we assume that the only pipes modeling blood vessels that are permeable are those that model capillaries.

The flow that passes through the walls of a permeable pipe is denoted by $J_i$. According to the common convention, $J_i>0$ denotes fluid leaving the pipe and $J_i<0$ fluid entering the pipe. Due to conservation of mass, the flow that leaves from an impermeable pipe $Q^{out}_i$ is the same as the flow that enters $Q^{in}_i$, thus:
\begin{equation}\label{eq:first}
Q^{out}_i=Q^{in}_i
\end{equation}
while the flow that leaves a permeable pipe is given by:
\begin{equation} \label{eq:1}
Q^{out}_i = Q^{in}_i - J_i
\end{equation}
We assume that the flow crossing through the walls of renal tubules and glomerular capillaries are constant fractions of the corresponding inflow:
\begin{equation}\label{eq:flux}
J_i=f_iQ_i^{in}
\end{equation}
where $f_i$ is the fraction of fluid that crosses through the pipe's wall. For the fractional coefficients $f_i$ we use the values listed on Table~\ref{tb:1}, which are chosen such that the model predicts flows similar to the antidiuretic rat model in~\cite{Moss:2014aa}.

Flow through the walls of the cortical and medullary capillaries are computed by the Starling equation \cite{Wolgast1981Functional-char}
\begin{align}
J_9&= K_f^{9}\ (P_9 - P_1 + \pi_9 - \pi_1) \label{eq:J9}
\\
J_{16} &= K_f^{16}\ (P_{16} - P_2 + \pi_{16} - \pi_2)\label{eq:J16}
\end{align}
where $K_f^9=1.59~\mu$m$^3$/mmHg/min and $K_f^{16}=2.28~\mu$m$^3$/mmHg/min are the filtration coefficients of the cortical and medullary capillaries and $\pi_1$, $\pi_2$, $\pi_9$, and $\pi_{16}$ are the oncotic pressures and $P_1$, $P_2$, $P_9$, and $P_{15}$ are the hydrostatic pressures in the associated compartments. The oncotic pressures are obtained by an approximation of the  Landis-Pappenheimer relation
\begin{equation}{\label{eq:oncotic}}
\pi_i = \alpha C_i+ \beta C_i^{2}
\end{equation}
where $\alpha = 1.63$~mmHg$\cdot$dl/gr and $\beta = 0.29$~mmHg$\cdot$dl$^2/$gr$^2$ as used in \cite{Deen1972A-model-of-glom}. In equation~\eqref{eq:oncotic}, $C_i$ denotes the concentration of protein in the compartment $i$. We assume a fixed protein concentration of the blood entering through the renal artery of $C_a= 5.5$~gr/dl and compute concentrations throughout the blood vessels (compartments 3--9 and 12--16) by taking into consideration conservation of mass
\begin{equation} {\label{eq:12}}
C^{out}_i = \frac{Q^{in}_i}{Q^{in}_i - J_i} C^{in}_i
\end{equation}
where $C_i^{in}$ and $C_i^{out}$ denote the inflow and outflow concentrations of the compartment $i$. The oncotic pressures $\pi_9$ and $\pi_{16}$ at equations~\eqref{eq:J9} and \eqref{eq:J16} are computed based on the averages
\begin{align}
C_9&=\frac{C_9^{in}+C_9^{out}}{2}
\\
C_{16}&=\frac{C_{19}^{in}+C_{19}^{out}}{2}
\end{align}

In each pipe and glomerulus, the internal pressure is denoted $P^{int}_i$ and the  external $P^{ext}_i$. For pipes, $P^{int}_i$ is computed by the average of the pressures at the associated inflow and outflow nodes (Figure~\ref{NephronModel}). For the glomeruli, internal pressure equals to the pressure of the associated node (Figure~\ref{NephronModel} and Table~\ref{tb:1}). For all pipes and glomerulus compartments, the external pressures equal the internal pressure of the surrounding compartment, which, in the case of the cortical and medullary regions, are denoted by $P_1$ and $P_2$, respectively. Exceptions to this are the arcuate arteries and veins (compartments 4 and 19, respectively), which anatomically are located between the cortex and the medulla \cite{Kriz1988A-standard-nome}, so we compute $P^{ext}_i$ for these compartments by the average of $P_1$ and $P_2$.

The volumes of the compartments, besides the regions and the afferent arterioles (compartments 1, 2 and 6, 12), depend \emph{passively} on the pressure difference that is developed across their walls:
\begin{equation} \label{eq:pressure_volume}
V_i = V^{ref}_i + s_i\ (P^{int}_i - P^{ext}_i + \Delta P^{ref}_i)
\end{equation}
where $V_i^{ref}$, $\Delta P^{ref}_i$, and $s_i$ are constants. In particular, $V_i^{ref}$ denotes a reference volume, and $\Delta P^{ref}_i$ denotes the pressure difference across the walls of the compartment when $V_i$ equals $V_i^{ref}$. The parameters $s_i$ are a measure of the distensibility of the compartments. A large $s_i$ value indicates a compartment that is very distensible, while a low value $s_i$ indicates a more rigid compartment. In the model, we use $s_i\ge0$ such that an increase in $P^{int}_i$ or a decrease in $P^{ext}_i$ leads to an expansion of the volume $V_i$, and {\it vise versa}.

For a model pipe, let $P^{in}_i$ and $P^{out}_i$ denote the pressures at its inflow and outflow nodes, respectively. These pressures are related by a modified form of the Poiseuille law:
\begin{equation} \label{eq:pois}
P^{in}_i - P^{out}_i = \frac{8 \mu_iL_i}{\pi R_i^{4}} \Big(Q^{in}_i - \frac{2}{3}J_i\Big)
\end{equation}
where $\mu_i$ is the viscosity of the flowing fluid, $L_i$ is the length of the pipe, and $R_i$ is its radius. In the model, we assume $\mu_i$ and $L_i$ to be constants, while we compute $R_i$ based on the compartment's volume (i.e. $V_i=\pi R_i^2L_i$). Equation~\eqref{eq:pois} reduces to the common Poiseuille equation for the impermeable pipes \cite{Sgouralis2015Mathematical-mo}, while for the permeable pipes, it is assumed that $J_i$ is linearly distributed along the length of the pipe with a value of 0 at the end of the pipe.

Pressure at node c1 equals the arterial blood pressure $P_a$, which in our model is a free variable. Pressures at nodes c18 and c32 are kept constant at 4~mmHg and 2~mmHg, respectively, in agreement with the values of venous and ureter pressures used in previous modeling studies \cite{Moss2013Hormonal-regula,Layton2012Signal-Transduc_B}.

\subsubsection{Model Afferent Arterioles}

The afferent arterioles are unique vessels in the sense that they \emph{actively} adjust radii such that blood flows through them at a fixed rate \cite{Holstein-Rathlou1994Renal-blood-flo,Sgouralis2015Mathematical-mo}. In the model, we assume that blood flow in the afferent arterioles that feed the short and long nephrons (i.e. $Q_6$ and $Q_{12}$, respectively) are kept fixed at 280~nl/min and 336~nl/min, respectively, as in previous modeling studies of renal hemodynamics, for example \cite{Moss:2014aa,Fry2014Impact-of-renal,Sgouralis2014Theoretical-Ass}.

We compute the radii of the afferent arterioles by the Poiseuille equation \cite{Sgouralis2015Mathematical-mo}, which yields
\begin{align}
R_6 &= \Big(\frac{8 \mu_6L_6}{\pi}\frac{Q_6}{P_{c4} - P_{c5}}\Big)^{1/4} \label{eq:pois_AA_1}
\\
R_{12} &= \Big(\frac{8 \mu_{12}L_{12}}{\pi}\frac{Q_{12}}{P_{c3} - P_{c10}}\Big)^{1/4} \label{eq:pois_AA_2}
\end{align}
Note that equations~\eqref{eq:pois_AA_1} and \eqref{eq:pois_AA_2} imply that whenever the pressure difference along the afferent arterioles $P_{c4}-P_{c5}$ and $P_{c3}-P_{c10}$ increases, the radii $R_6$ and $R_{12}$ decrease. This, in turn, implies that whenever the arterial blood pressure  $P_a$ increases, the afferent arterioles constrict, and thus the total volumes occupied by them, $V_6=\pi R_6^2L_6$ and $V_{12}=\pi R_{12}^2L_{12}$ are reduced.

\subsubsection{Model Interstitial Regions}

The cortical and medullary interstitial spaces, i.e.\ compartments 1 and 2, lie outside of the compartments 3--35 and therefore must be calculated separately using a different set of equations. We obtain the first of such relationships by assuming that the net accumulation of interstitial fluid within the cortex and medulla is zero. That is
\begin{align}\label{eq:J_cortex_total}
J_9+\frac{J_{22}}{80}+\frac{J_{26}}{80}+\frac{J_{28}}{160}+\frac{J_{32}}{160} =0
\\
\label{eq:J_medulla_total}
J_{16}+\frac{J_{23}}{500}+\frac{J_{29}}{1000}+\frac{J_{34}}{72000}=0
\end{align}
where the flows $J_i$ are weighted based on the total number of the compartments contained in the full model (Table~\ref{tb:1}).

Equations~\eqref{eq:J9} and \eqref{eq:J16} require the oncotic pressures $\pi_1$ and $\pi_2$, which in turn require the cortical and medullary protein concentrations $C_1$ and $C_2$ for equation~\eqref{eq:oncotic}. Protein concentrations in the cortical and medullary regions are computed assuming that the total mass of protein contained in each region, $M_1$ and $M_2$, respectively, remains constant. Thus,
\begin{align}\label{eq:c1}
C_1&=\frac{M_1}{V_1}
\\
\label{eq:c2}
C_2&=\frac{M_2}{V_2}
\end{align}
We use the values $M_1=1.93$~mgr and $M_2=1.25$~mgr, which are computed such that the resulting model predicts reference pressures in the renal cortex and medulla of $\sim$6~mmHg, similar to those estimated experimentally \cite{Cowley1997Role-of-the-ren}.

Cortical and medullary interstitial volumes $V_1$ and $V_2$ are assumed to change proportionally; thus,
\begin{align}
\frac{V_1}{V_2} = \kappa \label{eq:18}
\end{align}
where $\kappa$ is the proportionality constant. The combined volume of the interstitial regions $V_1+V_2$ is calculated based on the total volume of the kidney $V_0$ according to
\begin{equation}{\label{eq:16}}
V_1 + V_2 = V_0 - V_{cortex} - V_{medulla}
\end{equation}
where $V_{cortex}$ and $V_{medulla}$ are found by summing the total volumes of the pipe and glomerulus compartments contained within each region. Finally, the total volume of the kidney $V_0$ is calculated by
\begin{equation}\label{eq:capsule}
V_0 = V_0^{ref} + s_0 \ (P_1 - P^{ext}_0 + \Delta P_0^{ref})
\end{equation}
where in this case $P^{ext}_0$ refers to the pressure external to the kidney, which is set to 0~mmHg. Equation~\eqref{eq:capsule} assumes that the total volume of the kidney is determined by the distensibility of the renal capsule $s_0$, which is stretched by the difference of the pressures developed across it, i.e. $P_1-P^{ext}_0$.

\subsection{Model Parameters}
\label{sec:params}

Values for the model parameters are given in Table~\ref{tb:3}. These values are chosen such that at a reference arterial blood pressure $P_a^{ref}$ = 100~mmHg the model predicts pressures and volumes that are in good agreement with either direct experimental measurements \cite{Nyengaard1993Number-and-dime,Nordsletten2006Structural-morp,Jensen1977Angiotensin-II-,Heilmann2012Quantification-,Cortes1996Regulation-of-g} or previous modeling studies \cite{Moss:2014aa,Moss2013Hormonal-regula,Edwards2011Modulation-of-o,Sgouralis2012Autoregulation-,Oien1991A-multinephron-,Sgouralis2015Conduction-of-f,Sgouralis2015Renal-hemodynam,Sgouralis2013Control-and-Mod,Chen2011A-mathematical-}. 

The pressure-volume relationships used in the model, equations~\eqref{eq:pressure_volume} and \eqref{eq:capsule}, require values for the parameters $s_i$. We assume that (i) $s_i$ scale proportionally to the reference volumes
\begin{align}
s_i&=\sigma_i\ V_i^{ref}
\end{align}
and (ii) the coefficients $\sigma_i$ depend only on the histology of the associated compartment. That is, we group the compartments as follows:
\begin{itemize}
\item Group G1: renal capsule ($s_0$) and papillary collecting duct ($s_{35}$)
\item Group G2: glomeruli ($s_{21}$ and $s_{27}$)
\item Group G3: renal tubules ($s_{22}$--$s_{26}$) and proximal collecting ducts ($s_{28}$--$s_{34}$)
\item Group G4: pre-afferent arteriole blood vessels ($s_3$--$s_5$)
\item Group G5: post-afferent arteriole blood vessels ($s_7$--$s_{11}$ and $s_{13}$--$s_{20}$)
\end{itemize}
Then we assign the same flexibility value $\sigma_i$ to all members of each group (Table~\ref{tb:3}). With this formulation, the model compartments in each histological group experience the same fractional change in volume whenever they are challenged by the same pressure gradient $P_i^{int}-P^{ext}_i$.

The available experimental data do not permit an accurate estimate of the values of the flexibility parameters. For this reason, we treat the flexibilities of the five groups $\sigma_g$ as \emph{independent random variables}. To facilitate the comparison among the different groups, we set
\begin{equation} \label{eq:LAMBDA}
\sigma_g=\tilde \sigma_g\Lambda_g
\end{equation}
where $\tilde\sigma_g$ are constants, and $\Lambda_g$ are random variables configured to have mode 1. We estimate the values of $\tilde\sigma_g$ empirically based on \emph{ex vivo} measurements reported in~\cite{Hebert1975Whole-kidney-vo, Zhu1992Quantitative-re,Cortes1996Regulation-of-g,Cortell1973A-definition-of,Yamamoto1983Circulatory-pre} (Table~\ref{tb:3}).

\begin{figure}[tbp]
\begin{center}
	\includegraphics[scale=0.75]{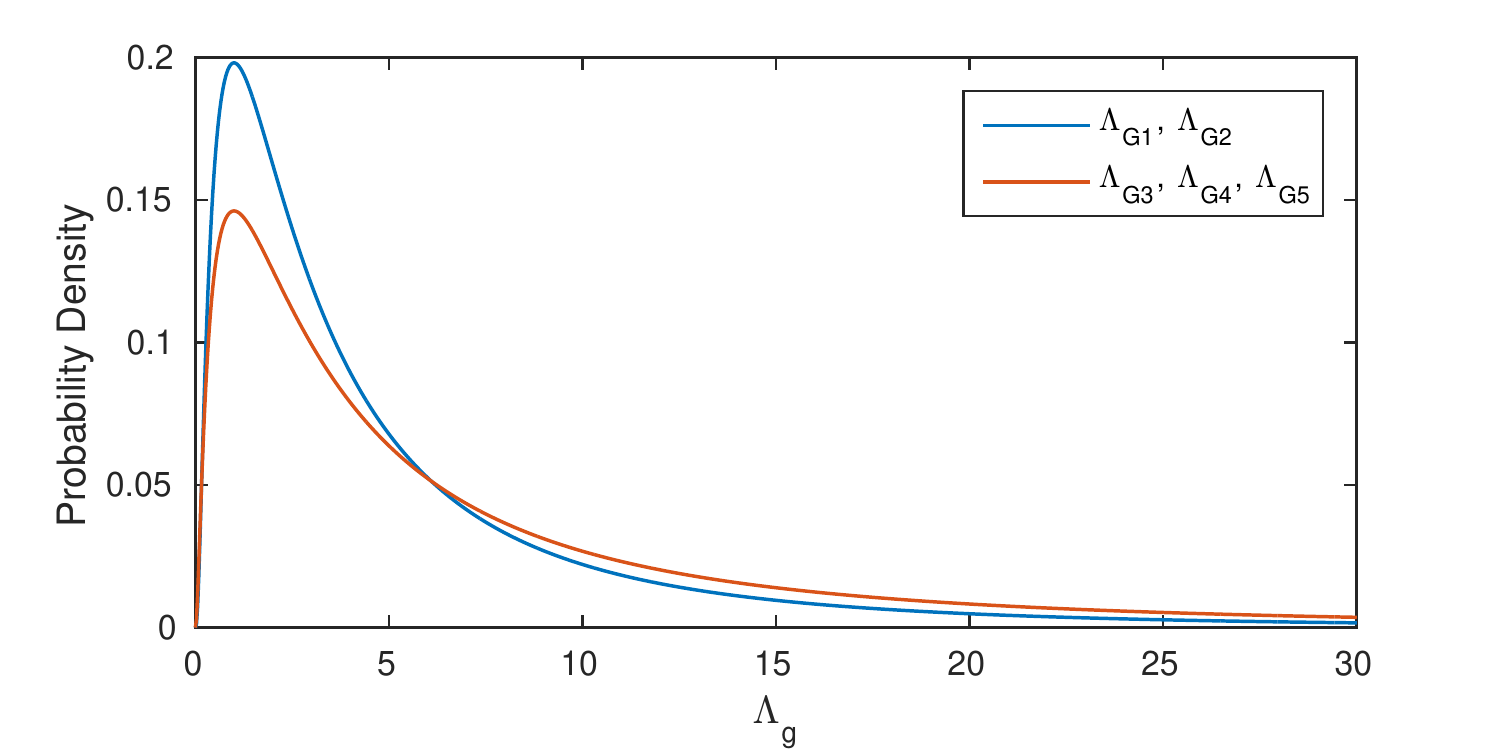}
	\caption{Probability densities of the flexibility parameters $\Lambda_g$ of the histological groups G1--G5 used in this study.}
	\label{fig:Lambda_pdfs}
\end{center}
\end{figure}

For each simulation, $\Lambda_g$ are drawn from the log-normal distribution (Figure~\ref{fig:Lambda_pdfs}), which is chosen such that (i) $s_g$ attain non-negative values, (ii) arbitrarily large values of $s_g$ are allowed, and (iii) low $s_g$ values are more frequent than large ones. We choose the latter condition assuming that the experimental procedures (anesthesia, renal decapsulation, tissue isolation, etc.) utilized in \cite{Hebert1975Whole-kidney-vo,Cortes1996Regulation-of-g,Cortell1973A-definition-of,Yamamoto1983Circulatory-pre} likely increase rather than decrease tissue flexibility, thus our computed $\tilde\sigma_g$ likely overestimate rather than underestimate $\sigma_g$.

Finally, we configure the log-normal distributions such that $\Lambda_{G1}$ and $\Lambda_{G2}$ have log-standard deviation of 1.1, and $\Lambda_{G3}$, $\Lambda_{G4}$, and $\Lambda_{G5}$ have log-standard deviation of 1.25 (Figure~\ref{fig:Lambda_pdfs}). According to our experience, such configuration reflects the degree of the uncertainty in our estimated values of $\tilde\sigma_g$, for which we consider $\tilde\sigma_{G3}$, $\tilde\sigma_{G4}$, and $\tilde\sigma_{G5}$ less accurately estimated than $\tilde\sigma_{G1}$ and $\tilde\sigma_{G2}$.

\subsection{Sensitivity Analysis}
\label{sec:sens}

\subsubsection{Formulation}

For the sensitivity analysis of the model described in the previous sections, we adopt a \emph{variance-based method} which is best suited for non-linear models  \cite{Saltelli2000Sensitivity-ana,Sobol2001Global-sensitiv}. Let
\begin{equation}
y=f(x_1,x_2,\dots,x_k)
\end{equation}
denote a generic model, where $y$ is an output value and $x_1,x_2\dots,x_k$ are some random inputs (in our case those represent the uncertain parameters). For a factor $x_g$, the first- and total-order sensitivity indices are given by
\begin{gather}
S_g=\frac{\mathbb{V}(\mathbb{E}(y|x_g))}{\mathbb{V}(y)}	\label{eq:Sj}
\\
T_g=1-\frac{\mathbb{V}(\mathbb{E}(y|x_{-g}))}{\mathbb{V}(y)}	\label{eq:Tj}
\end{gather}
respectively, \cite{Saltelli2000Sensitivity-ana,Saltelli2002Making-best-use,Sobol2001Global-sensitiv}. In the equations above, $\mathbb{E}$ and $\mathbb{V}$ denote mean value and variance, respectively. In \eqref{eq:Sj}, first the mean of $y$ is computed by fixing the factor $x_g$ to some value $\tilde x_g$, and then the variance of the mean values is computed over all possible $\tilde x_g$. In~\eqref{eq:Tj}, first the mean value is computed by fixing all factors except $x_g$ (which is denoted by $x_{-g}$), and then the variance of the mean values is computed over all possible $x_{-g}$.

According to the above definitions, the first-order index $S_g$ indicates the fraction by which the variance of $y$ will be reduced if only the value of the factor $x_g$ is certainly specified \cite{Saltelli2000Sensitivity-ana}. Similar, the total-order index $T_g$ indicates the fraction of the variance of $y$ that will be left if all factors besides $x_g$ are certainly specified \cite{Saltelli2000Sensitivity-ana}. We compute both indices, because generally for a non-linear model the factors are expected to interact in a non-additive way, and therefore $T_g$ is expected to be larger than $S_g$. The difference $T_g-S_g$ characterizes the extent of the interactions with the other factors that $x_g$ is involved with.

\subsubsection{Evaluation of Sensitivity Indices}

To better characterize the contribution of the individual factors $\Lambda_g$ of equation~\eqref{eq:LAMBDA}, in the variance of $P_1$ and $P_2$, we calculate their first- and total-order sensitivity indices given in~\eqref{eq:Sj} and \eqref{eq:Tj}. We compute the indices according to the method proposed by Saltelli \cite{Saltelli2002Making-best-use}, which is computationally less demanding than a straightforward application of the formulas in \eqref{eq:Sj} and \eqref{eq:Tj}.

Briefly, according to the Saltelli method we form two input matrices:
\begin{gather}
M_A=
\begin{bmatrix}
\Lambda_{G1}^{1,A}	&\Lambda_{G2}^{1,A}	&\Lambda_{G3}^{1,A}	&\Lambda_{G4}^{1,A}	&\Lambda_{G5}^{1,A}	&\Lambda_{G6}^{1,A}	\\
\Lambda_{G1}^{2,A}	&\Lambda_{G2}^{2,A}	&\Lambda_{G3}^{2,A}	&\Lambda_{G4}^{2,A}	&\Lambda_{G5}^{2,A}	&\Lambda_{G6}^{2,A}	\\
\vdots			&\vdots				&\vdots				&\vdots				&\vdots				&\vdots				\\
\Lambda_{G1}^{N,A}	&\Lambda_{G2}^{N,A}	&\Lambda_{G3}^{N,A}	&\Lambda_{G4}^{N,A}	&\Lambda_{G5}^{N,A}	&\Lambda_{G6}^{N,A}	\\
\end{bmatrix}
\\
M_B=
\begin{bmatrix}
\Lambda_{G1}^{1,B}	&\Lambda_{G2}^{1,B}	&\Lambda_{G3}^{1,B}	&\Lambda_{G4}^{1,B}	&\Lambda_{G5}^{1,B}	&\Lambda_{G6}^{1,B}	\\
\Lambda_{G1}^{2,B}	&\Lambda_{G2}^{2,B}	&\Lambda_{G3}^{2,B}	&\Lambda_{G4}^{2,B}	&\Lambda_{G5}^{2,B}	&\Lambda_{G6}^{2,B}	\\
\vdots			&\vdots				&\vdots				&\vdots				&\vdots				&\vdots				\\
\Lambda_{G1}^{N,B}	&\Lambda_{G2}^{N,B}	&\Lambda_{G3}^{N,B}	&\Lambda_{G4}^{N,B}	&\Lambda_{G5}^{N,B}	&\Lambda_{G6}^{N,B}	\\
\end{bmatrix}
\end{gather}
by generating Monte Carlo samples $\Lambda_{g}^{j,A}$ and $\Lambda_{g}^{j,B}$ for the factors $\Lambda_g$. Subsequently, for each factor, we forme a matrix $M_g$. Each $M_g$ is formed by the columns of $M_A$, except the column that corresponds to the factor $\Lambda_g$, which is taken from $M_B$. For instance, $M_{G2}$ is given by:
\begin{gather}
M_{G2}=
\begin{bmatrix}
\Lambda_{G1}^{1,A}	&\mathbf{\Lambda_{G2}^{1,B}}	&\Lambda_{G3}^{1,A}	&\Lambda_{G4}^{1,A}	&\Lambda_{G5}^{1,A}	&\Lambda_{G6}^{1,A}	\\
\Lambda_{G1}^{2,A}	&\mathbf{\Lambda_{G2}^{2,B}}	&\Lambda_{G3}^{2,A}	&\Lambda_{G4}^{2,A}	&\Lambda_{G5}^{2,A}	&\Lambda_{G6}^{2,A}	\\
\vdots			&\mathbf{\vdots	}			&\vdots				&\vdots				&\vdots				&\vdots				\\
\Lambda_{G1}^{N,A}	&\mathbf{\Lambda_{G2}^{N,B}}	&\Lambda_{G3}^{N,A}	&\Lambda_{G4}^{N,A}	&\Lambda_{G5}^{N,A}	&\Lambda_{G6}^{N,A}	\\
\end{bmatrix}
\end{gather}
We use each row of the matrices $M_A$, $M_B$, and $M_g$ to solve the model equations at $P_a=180$~mmHg and combine the solutions in the vectors:
\begin{align}
m_A^k=
\begin{bmatrix}
P_k^{1,A}	\\
P_k^{2,A}	\\
\vdots	\\
P_k^{N,A}	\\
\end{bmatrix},
\qquad
m_B^k=
\begin{bmatrix}
P_k^{1,B}	\\
P_k^{2,B}	\\
\vdots		\\
P_k^{N,B}	\\
\end{bmatrix},
\qquad
m_g^k=
\begin{bmatrix}
P_k^{1,g}	\\
P_k^{2,g}	\\
\vdots		\\
P_k^{N,g}	\\
\end{bmatrix}
\end{align}
where $k=1$ corresponds to the pressure in the cortical region $P_1$, and $k=2$ to the pressure in the medullary region $P_2$. The first- and total-order sensitivity indices are then computed by
\begin{equation}{\label{eq:30}}
S^k_g=\dfrac{\frac{1}{N-1}\sum_{j=1}^N\big(P_k^{j,A}P_k^{j,g}\big)-\frac{1}{N}\sum_{j=1}^N\big(P_k^{j,A}P_k^{j,B}\big)}
{\mathbb{V}\big(m_A^k\big)}
\end{equation}
\begin{equation}{\label{eq:31}}
T^k_g=1-\dfrac{\frac{1}{N-1}\sum_{j=1}^N\big(P_k^{j,B}P_k^{j,g}\big)-\big(\frac{1}{N}\sum_{j=1}^NP_k^{j,B}\big)^2}
{\mathbb{V}\big(m_B^k\big)}
\end{equation}
respectively. In the above equations (\ref{eq:30}) - (\ref{eq:31}) , $\mathbb{V}$ denotes the sample variance. For further details on the method, see \cite{Saltelli2002Making-best-use}.

\subsection{Numerical Methods}

For the numerical solution, we combine the model equations \eqref{eq:first}--\eqref{eq:capsule} into a system of 69 coupled non-linear equations. Given a value for the arterial blood pressure $P_a$ and a choice for the flexibility parameters $\Lambda_g$, the resulting system is solved to yield the values for the pressures at the interstitial regions $P_1$ and $P_2$, the pressures at the model nodes $P_{c1}$--$P_{c32}$, and the volumes of the compartments $V_1$--$V_{35}$.

To obtain solutions, we implement the system in \texttt{MATLAB} and use the standard root-finding function (\texttt{fsolve}).
This function computes solutions to the model equations iteratively by starting from a given initial approximation. For the initial approximation we use the reference values from literature (Table~\ref{tb:3}). Note that by the construction of the model, the solution at reference can be obtained trivially, and thus no root-finding is necessary for this step.

%%%%%%%%%%%%%%%%%%%%%%%%%%%%%%%%%%%%%%%%%%%%%%%%%%%%%%%%%%%%%%%%
%%%%%%%%%%%%%%%%%%%%%%%%%%%%%%%%%%%%%%%%%%%%%%%%%%%%%%%%%%%%%%%%
\section{Results}

\subsection {Selected Case Studies}
\label{sec:select}

\begin{figure}[tb]
\begin{center}
	\includegraphics[scale=0.75]{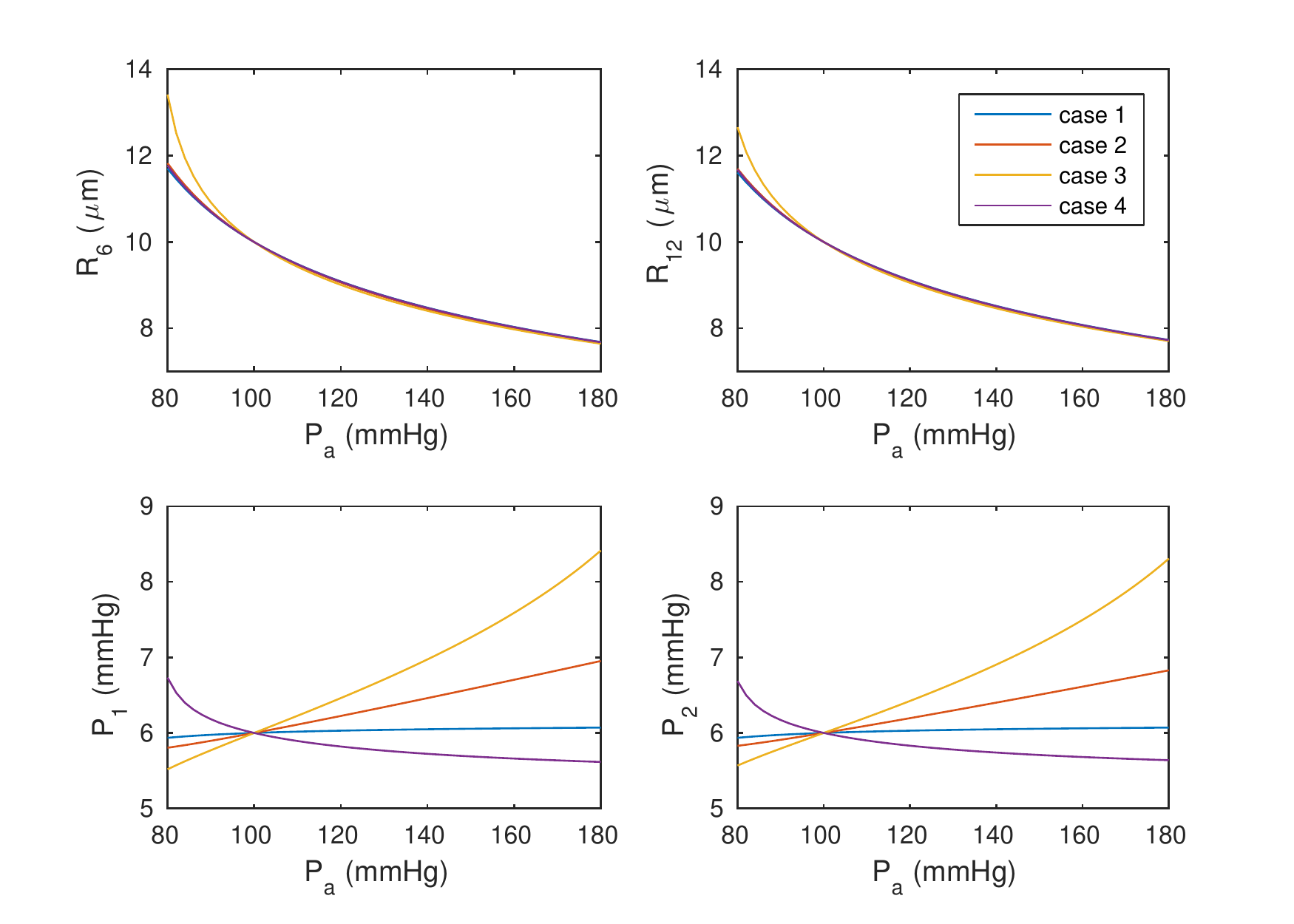}
	\caption{Model predictions for selected parameter choices. \emph{Upper panels:} radii of the afferent arterioles. \emph{Lower panels:} pressures in the interstitial regions.}
	\label{fig:selected}
\end{center}
\end{figure}

In the first set of simulations, we investigate how the pressures in the interstitial regions $P_1$ and $P_2$ are affected by the arterial blood pressure $P_a$ for selected choices of the flexibility parameters when $P_a$ varies in the range 80--180~mmHg. In particular, we make the following choices for the flexibility parameters:
\begin{itemize}
\item Case 1: $\Lambda_{G1}=\Lambda_{G2}=\Lambda_{G3}=\Lambda_{G4}=\Lambda_{G5}=0$
\item Case 2: $\Lambda_{G1}=\Lambda_{G2}=\Lambda_{G3}=\Lambda_{G4}=\Lambda_{G5}=1$
\item Case 3: $\Lambda_{G1}=\Lambda_{G2}=\Lambda_{G3}=\Lambda_{G4}=\Lambda_{G5}=4$
\item Case 4: $\Lambda_{G1}=2.7$, $\Lambda_{G2}=0.04$, $\Lambda_{G3}=0$, $\Lambda_{G4}=0.27$, $\Lambda_{G5}=0.2$
\end{itemize}
Figure~\ref{fig:selected} shows key solution values.

Case 1 corresponds to a kidney with rigid compartments. In this case, pressure does not affect the volume of the compartments except of the two afferent arterioles $V_6$ and $V_{12}$. For example, at elevated $P_a$, the pressure differences along the afferent arterioles $P_{c4}-P_{c5}$ and $P_{c3}-P_{c10}$ increase. As a result, the arterioles constrict in order to maintain constant blood flow (equations~\eqref{eq:pois_AA_1} and \eqref{eq:pois_AA_2}). Given that total kidney volume $V_0$ does not change as given by equation~\eqref{eq:capsule}, the reduction in afferent arteriole volume increases the volume of the interstitial regions $V_1$ and $V_2$ given by equation~\eqref{eq:18}. In turn, increases in  interstitial volumes reduce the protein concentrations $C_1$ and $C_2$ by equations~\eqref{eq:c1}--\eqref{eq:c2} and the oncotic pressures $\pi_1$ and $\pi_2$ that promote uptake $J_9$ and $J_{16}$ of interstitial fluid by equations~\eqref{eq:J9}--\eqref{eq:J16}. However, due to tubular reabsorption $J_{22}$--$J_{34}$, the flow of fluid into the interstitial spaces is kept constant (equations~\eqref{eq:J_cortex_total} and \eqref{eq:J_medulla_total}). Thus, in order to maintain a constant uptake and avoid accumulation of interstitial fluid, $P_1$ and $P_2$ increase. \emph{Vise versa}, a decrease in $P_a$ has the opposite effects and results in a decrease of $P_1$ and $P_2$. Because the total volume of the afferent arterioles is only a minor fraction of the volume of the interstitial regions ($\sim$2\%, see Table~\ref{tb:3}), even large changes of $R_6$ and $R_{12}$ induce small changes of $\pi_1$ and $\pi_2$. Therefore, the total change in $P_1$ and $P_2$, across the full range of $P_a$ variation, is in the order of 0.1~mmHg (see blue curves in Figure~\ref{fig:selected}).

Case 2 corresponds to a kidney with distensible compartments. This case is similar to case 1; however, the changes of $P_1$ induced by the constriction of the afferent arterioles is followed by an expansion of the renal capsule (equation~\eqref{eq:capsule}), which increases whole kidney volume $V_0$. So, in this case, the cortical and medullary interstitial volumes $V_1$ and $V_2$ increase to a larger extent compared with case 1 in order to accommodate the expansion of $V_0$. As a result, interstitial protein concentrations $C_1$, $C_2$, and oncotic pressures $\pi_1$, and $\pi_2$ drop by larger amounts than in case 1. Consequently, significant drops in $P_1$ and $P_2$ follow (see orange curves in Figure~\ref{fig:selected}).

Case 3 corresponds to a kidney with very flexible compartments and renal capsule. Through the same effects as in cases 1 and 2, changes in arterial pressure $P_a$ lead to similar changes in $P_1$ and $P_2$. Because in this case the expansion of whole kidney volume $V_0$ is greater than in case 2, due to the increased flexibility of the renal capsule $s_0$, the interstitial pressures are affected to a greater extent too (see yellow curves in Figure~\ref{fig:selected}).

Case 4 shows a different behavior that corresponds to a kidney with flexible capsule but relatively rigid compartments. As in all cases, $P_a$ affects severely the pressures in the pre-afferent arteriole vascular compartments $P_3$, $P_4$, and $P_5$ (equation \eqref{eq:pois}), which are not regulated by the active constriction/dilation of the afferent arterioles. As a result, whenever $P_a$ increases, $P_3$, $P_4$, and $P_5$ also increase, leading to an increase of the associated pre-afferent arteriole vascular volumes $V_3$, $V_4$, and $V_5$. Note that the increase of $V_3$, $V_4$, and $V_5$ opposes the reduction of $V_6$ and $V_{12}$ caused by constriction of the afferent arterioles. In this particular case, opposite to what happens in cases 1-3, the increase of the total volume of the pre-afferent arteriole compartments $V_3$, $V_4$, and $V_5$ exceeds the reduction of the total volume of the afferent arterioles $V_6$ and $V_{12}$. As a result, the interstitial regions are compressed, which in turn leads to increases of the protein concentrations $C_1$ and $C_2$ and oncotic pressures $\pi_1$ and $\pi_2$. Because the uptake of interstitial fluid is maintained constant, this leads to reductions of $P_1$ and $P_2$. Finally, the reductions of $P_1$ and $P_2$ are further amplified by constriction of the renal capsule that follows the reduction of $P_1$.

\subsection{Sensitivity Analysis}

From the previous section, it is apparent that the predictions of the model depend on the choice of the flexibility parameters $\Lambda_g$, which are not well-characterized (Section~\ref{sec:params}). To assess the degree to which different choices affect the pressures in the interstitial regions $P_1$ and $P_2$, we sample the parameter space. For each sample point, we evaluate the model solution at an elevated arterial blood pressure $P_a$. For all simulations, we keep $P_a$ constant at 180~mmHg.

\subsubsection{Summary Statistics}

\begin{figure}[t!b]
\begin{center}
	\includegraphics[scale=0.75]{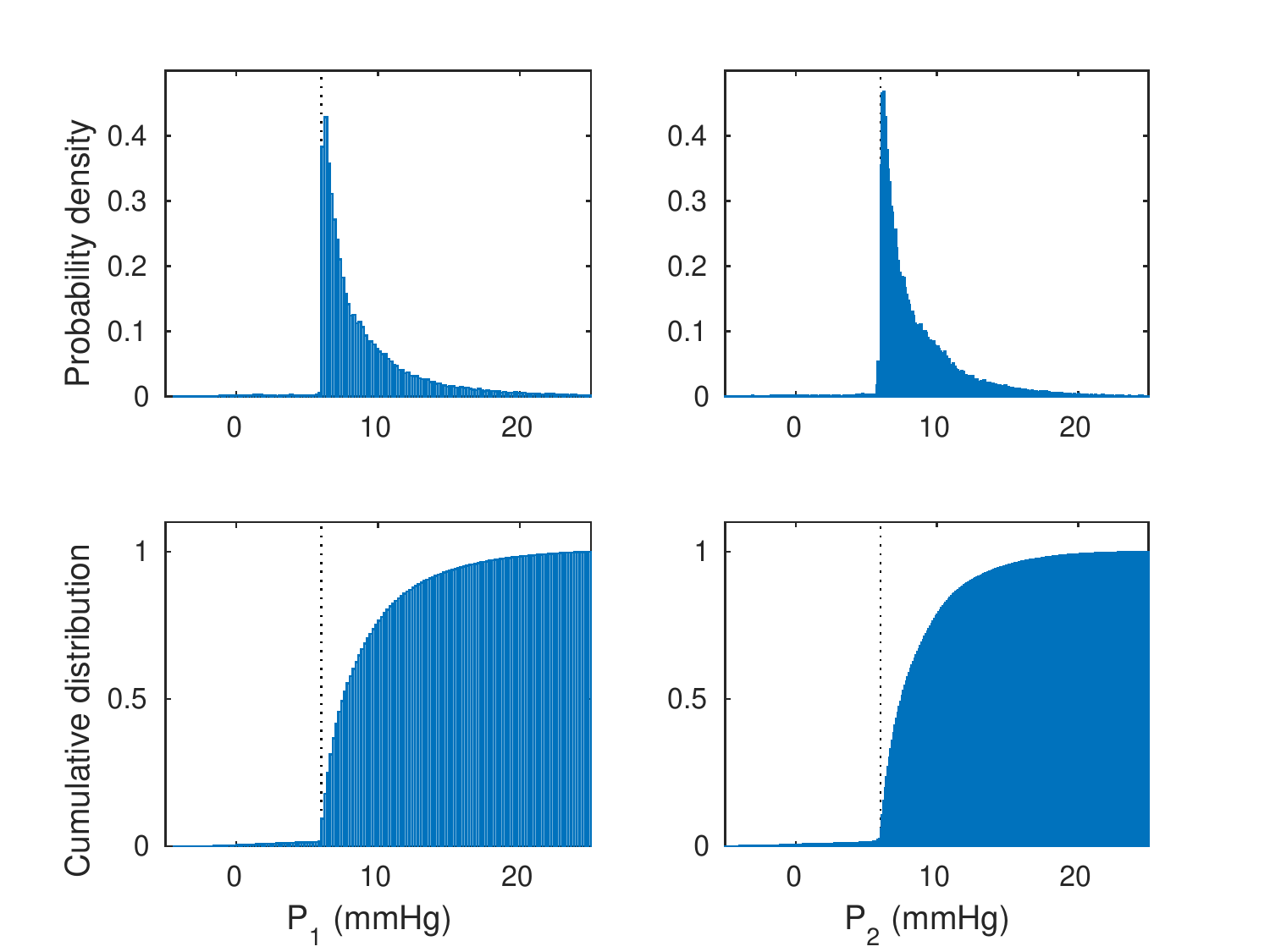}
	\caption{Probability densities of $P_1$ (left panel) and $P_2$ (right panel) at elevated arterial blood pressure ($P_a=180$~mmHg) as estimated by model simulations. Vertical lines indicate the values at the reference arterial blood pressure ($P_a=100$~mmHg).}
	\label{fig:hist_7_factors}
\end{center}
\end{figure}

The model utilizes 5 factors that correspond to the flexibility parameters associated with the histological groups of Section~\ref{sec:params}. We use a sample size of $N=41\times10^3$ 
and perform sampling with the Monte Carlo method. The resulting probability densities and cumulative distributions of $P_1$ and $P_2$ are shown in Figure~\ref{fig:hist_7_factors}.

As can be seen in Figure~\ref{fig:hist_7_factors}, the model predicts mostly increased $P_1$ and $P_2$ at elevated $P_a$. However, the uncertainty in the flexibility parameters $\Lambda_g$ induces a significant degree of variability for both pressures. The mean values of $P_1$ and $P_2$ are 9.1 and 8.6~mmHg, and the standard deviations are 4.1 and 3.7~mmHg, respectively. Both pressure distributions are heavily skewed towards large values.

Interestingly, the model also predicts low or even negative pressures. Negative pressure values indicate that the pressures in the interstitial regions fall below the pressure in the space surrounding the kidney $P_0^{ext}$, which in this study is set to 0~mmHg. In summary, 84\% of $P_1$ and 77\% of $P_2$ values at $P_a=180$~mmHg are above the corresponding values at $P_a=100$~mmHg, and 16\% of $P_1$ and 11\% of $P_2$ values lie below 0~mmHg or above 15~mmHg.

\begin{figure}[tb]
\begin{center}
	\includegraphics[scale=0.75]{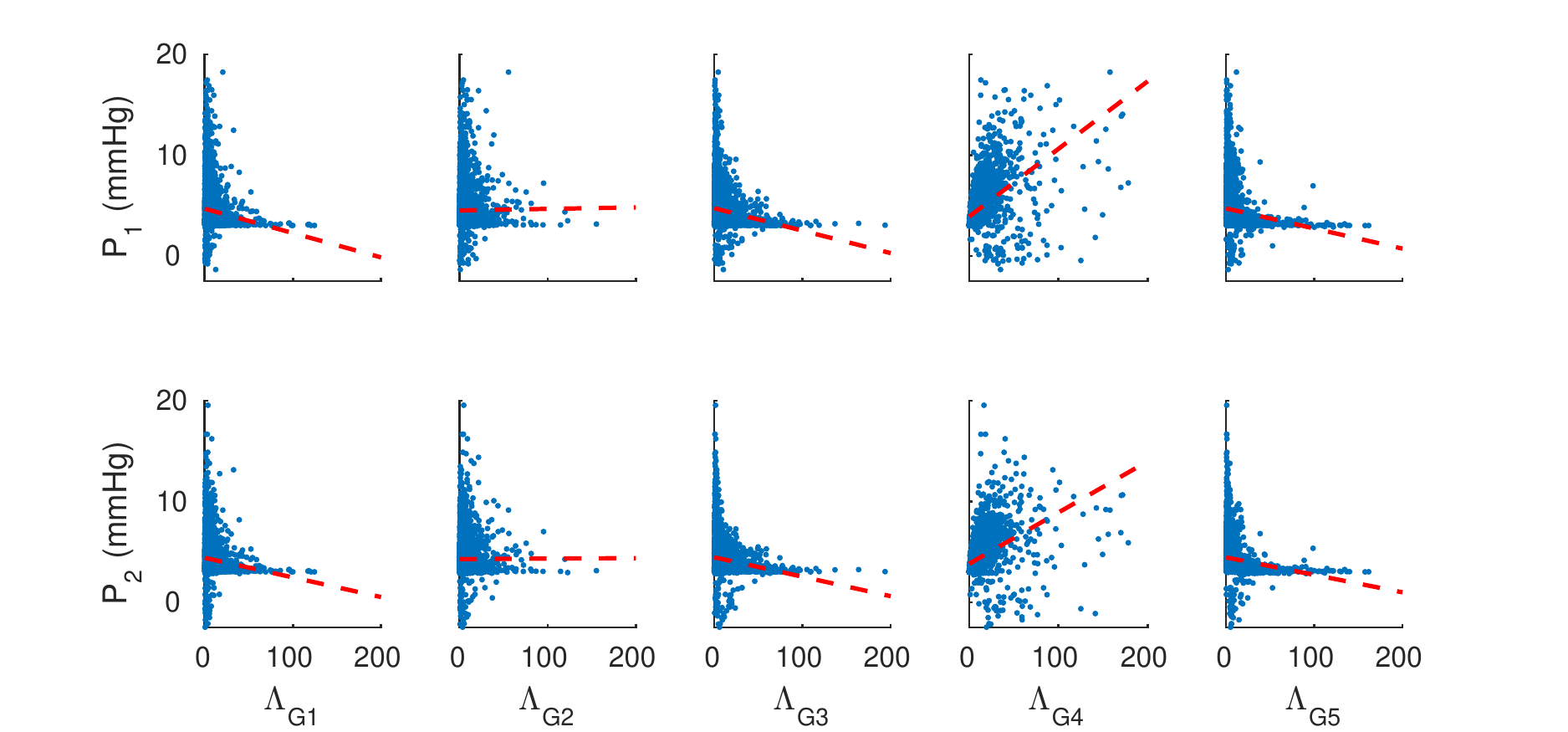}
	\caption{Interstitial pressures $P_1$ (upper panels) and $P_2$ (lower panels) with respect to the sampled input factors $\Lambda_g$. Dashed lines indicate the linear regression estimates. For clarity, only 1/5 of the computed points are shown.}
	\label{fig:scatter_7_factors}
\end{center}
\end{figure}

Scatter plots between the input factors $\Lambda_g$ and the computed pressures $P_1$ and $P_2$ are shown in Figure~\ref{fig:scatter_7_factors}. Only $\Lambda_{G4}$ shows a clear influence on $P_1$ and $P_2$, with high values of $\Lambda_{G4}$ being associated generally with higher interstitial pressures. No apparent trend can be identified for the rest of the factors. Linear regressions between the computed pressures and the input factors (shown by the dashed lines in Figure~\ref{fig:scatter_7_factors}) yield low $R^2$. Precisely, $R^2$ for $\Lambda_{G4}$ equal 0.25 for $P_1$ and 0.16 for $P_2$. The rest of the factors yield $R^2$ for 0.02 or less. Such low $R^2$ indicate strong non-linear dependencies of the interstitial pressures on the input factors, a behavior that most likely stems from the inverse-forth-power in the Poiseuille law given by equation~\eqref{eq:pois}. 

\begin{figure}[tb]
\begin{center}
	\includegraphics[scale=0.75,trim=0em 0em 0em 3ex]{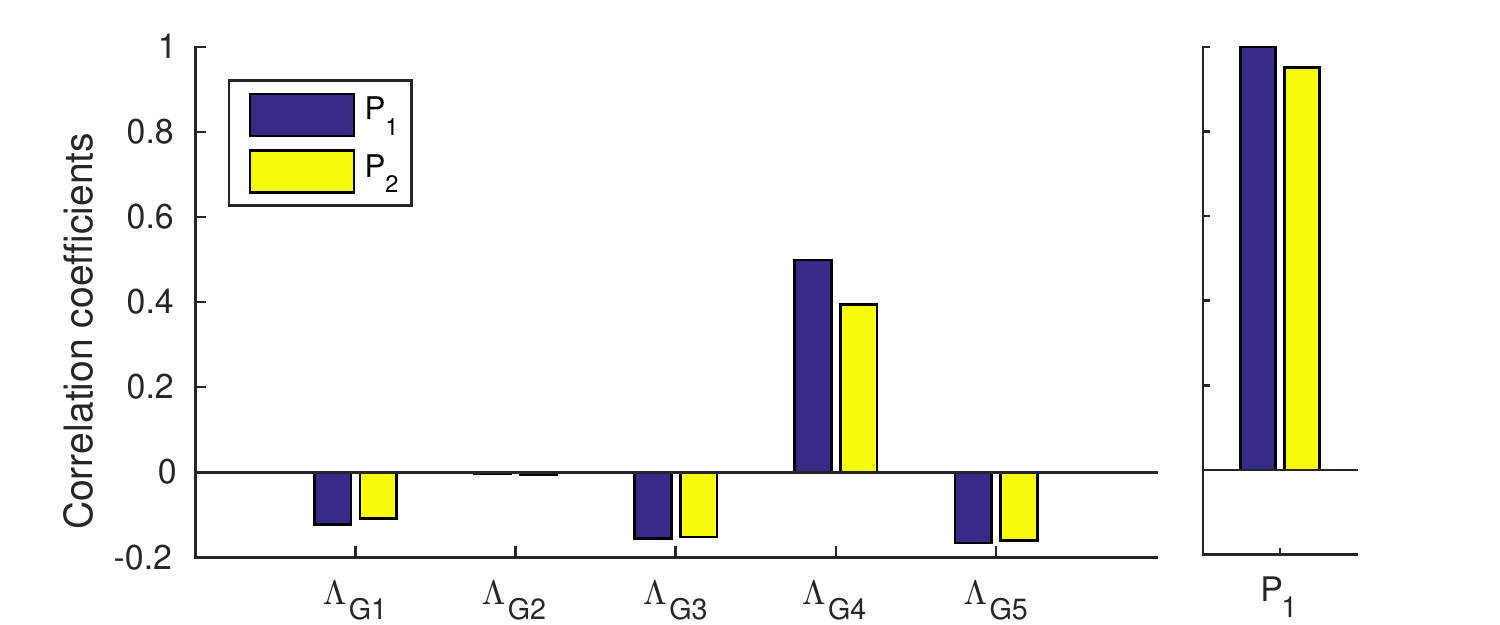}
	\caption{Correlation coefficients between the input factors $\Lambda_g$ and the computed pressures in the cortical and medullary interstitial spaces $P_1$ and $P_2$, respectively.}
	\label{fig:corr_7_factors}
\end{center}
\end{figure}

Correlation coefficients computed between the input factors $\Lambda_g$ and the computed pressures $P_1$ and $P_2$ are shown on Figure~\ref{fig:corr_7_factors} (left panel). As is suggested by Figure~\ref{fig:scatter_7_factors}, $\Lambda_{G4}$ is positively correlated, weakly though, with $P_1$ and $P_2$. From the rest of the factors, $\Lambda_{G1}$, $\Lambda_{G3}$, and $\Lambda_{G5}$ are negatively correlated with $P_1$ and $P_2$, however to an even weaker than for $\Lambda_{G4}$, and $\Lambda_{G2}$ shows no correlation with either $P_1$ or $P_2$.

In contrast to the apparent lack of any trend between the computed pressures $P_1$ and $P_2$ and the input factors $\Lambda_{G4}$, the model predicts a high degree of correlation between $P_1$ and $P_2$. The associated correlation coefficient reaches as high as 0.95 (Figure~\ref{fig:corr_7_factors} right panel), which indicates that $P_1$ and $P_2$ are predicted to change {\it in tandem} in a seemingly linear way.

\subsubsection{Sensitivity Indices}

To better characterize the contribution of the individual factors $\Lambda_g$ in the variance of $P_1$ and $P_2$, we calculate their first- and total-order sensitivity indices shown on equations~\eqref{eq:Sj} and \eqref{eq:Tj}. Details on the adopted computational methods can be found in Section~\ref{sec:sens}.

\begin{figure}[tb]
\begin{center}
	\includegraphics[scale=0.75]{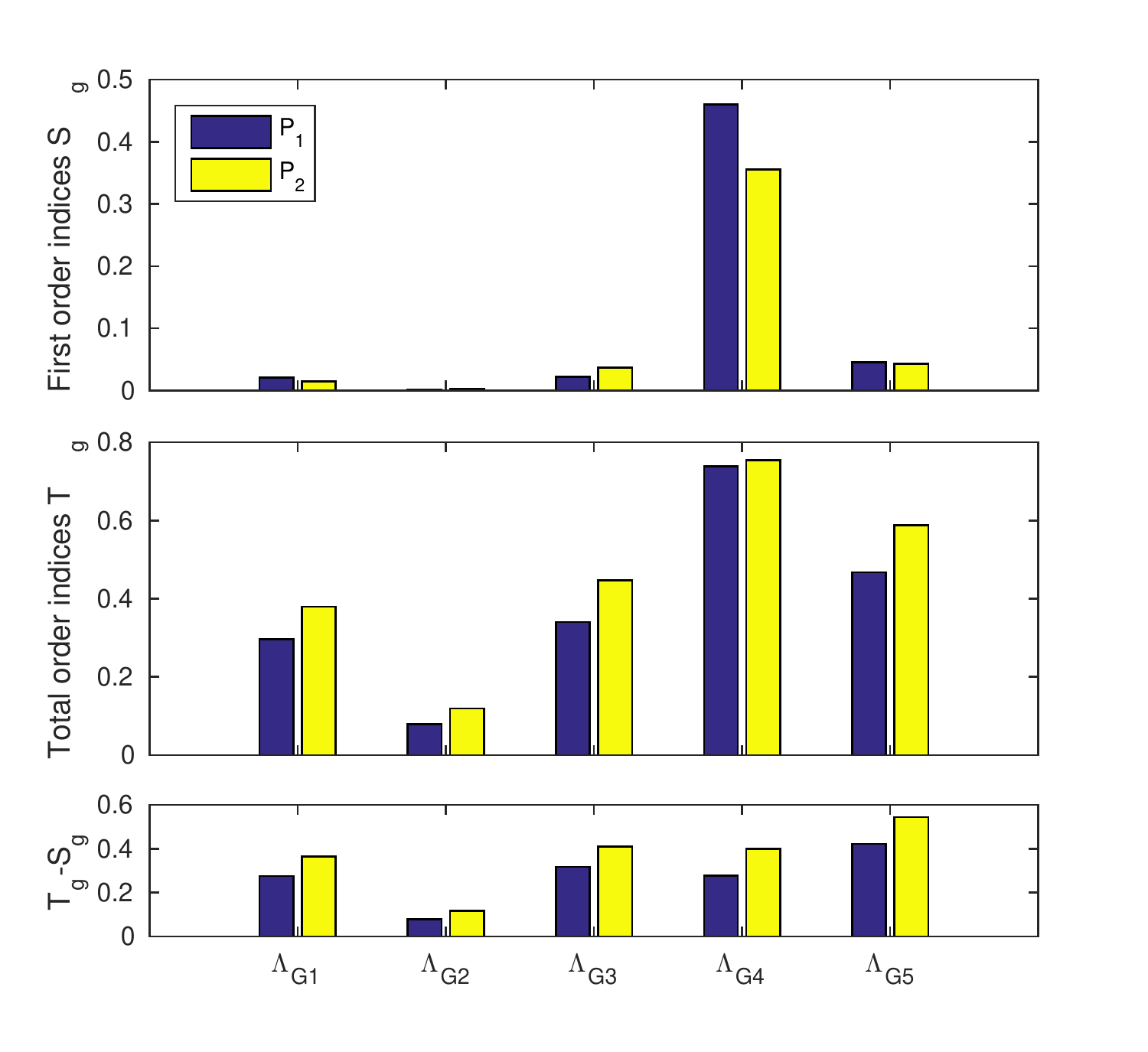}
	\caption{First-order (upper panel) and total-order (middle panel) sensitivity indices of $P_1$ and $P_2$ at elevated arterial blood pressure ($P_a=180$~mmHg). Lower panel shows the difference between the first- and total-order sensitivity indices.}
	\label{fig:indices_7_factors}
\end{center}
\end{figure}

Figure~\ref{fig:indices_7_factors} shows the computed indices. Evidently, the flexibility of the pre-afferent arteriole vascular segments (group G4) accounts for most of the variation in $P_1$ or $P_2$ with respect to either the first- or total-order indices. The post-afferent arteriole vasculature (group G5) has the second most significant contribution. Groups G1--G3 have only minor contributions according to the first-order sensitivity indices. However, this is not the case with the total-order indices, which indicate that G1 and G3 are involved to a significant degree in interactions. On the contrary, the glomeruli (group G2) have only a minor involvement in interactions.

For all groups, it is observed $T_g^1<T_g^2$ and $T_g^1-S_g^1<T_g^2-S_g^2$, which indicate that the medullary pressure $P_2$ is more susceptible to interactions than cortical pressure $P_1$. This behavior is expected, given that the afferent arterioles (compartments 6 and 12), which initiate the changes in $P_1$ and $P_2$, are located exclusively in the cortex, while the medulla is susceptible mostly to secondary interactions initiated by the expansion/constriction of the renal capsule.

%%%%%%%%%%%%%%%%%%%%%%%%%%%%%%%%%%%%%%%%%%%%%%%%%%%%%%%%%%%%%%%%
%%%%%%%%%%%%%%%%%%%%%%%%%%%%%%%%%%%%%%%%%%%%%%%%%%%%%%%%%%%%%%%%
\section{Conclusions}

We develop a multi-compartmental computational model of the rat kidney. The model is constructed using conservation laws (equations~\eqref{eq:1} and \eqref{eq:12}), fluid dynamics (equation~\eqref{eq:pois}), simplified pressure-volume relationships (equations~\eqref{eq:pressure_volume} and \eqref{eq:capsule}), and constitutive equations specific to the physiology of the kidney (equations~\eqref{eq:flux} and \eqref{eq:J_cortex_total}--\eqref{eq:J_medulla_total}).

We assign values to the model parameters (Tables~\ref{tb:1} and \ref{tb:3}) using experimental measurements when such measurements were available and previous modeling studies when direct measurements were not available. However, the data required for the flexibility parameters $\sigma_i$ are sparse and do not suffice for an accurate estimation of their values. To that end, we choose to model these parameters as random variables with probability distributions that permit values spanning multiple orders of magnitude (Section~\ref{sec:params} and Figure~\ref{fig:Lambda_pdfs}).

To determine the probability distributions of the random variables, we define five histological groups within the model kidney. \emph{Group G1} models thick and relatively inflexible structures, for which we use pressure-mass data obtained from whole kidneys in dogs \cite{Hebert1975Whole-kidney-vo,Zhu1992Quantitative-re}. \emph{Group G2} models the glomeruli, for which we use pressure-volume data measured in rats \cite{Cortes1996Regulation-of-g}. \emph{Group G3} models the various segments of the nephrons and the proximal parts of the collecting duct, for which we use pressure-radius measurements of the rat proximal tubule \cite{Cortell1973A-definition-of}. \emph{Groups G4 and G5} model the blood vessels, for which we use pressure-volume measurements of the systemic circulation measured in rats \cite{Yamamoto1983Circulatory-pre}. We combine the post-afferent arteriole vasculature in one group (group G5), despite that it consists of segments of the arterial and venous vascular trees \cite{Kriz1988A-standard-nome}. We are motivated to do so by the fact that these vascular segments have considerably thiner walls and therefore should be considerably more flexible than the pre-afferent arteriole segments \cite{rhodin2011architecture}.

Output from the model leads to a range of predictions depending on the choices of the flexibility values. Generally, increased arterial blood pressure is predicted to increase the pressure in both interstitial spaces (Figure~\ref{fig:hist_7_factors}). As arterial blood pressure increases from 100~mmHg to 180~mmHg, interstitial pressures are predicted to increase on average by $\sim$3~mmHg. Changes of similar magnitude have been observed in the kidneys of rats \cite{Garcia-Estan1989Role-of-renal-i,Khraibi2001Role-of-gender-,Skarlatos1994Spontaneous-cha,Khraibi2000Renal-interstit} and dogs \cite{Majid2001Nitric-oxide-de,Granger1988Effects-of-rena}. Upon a limited number of flexibility choices, however, the model predicts decreased interstitial pressures as a result. Further, the model predicts a tight correlation between the cortical and the medullary pressures, Figure~\ref{fig:corr_7_factors} (right panel), which is also in agreement with the experimental observations reported in \cite{Garcia-Estan1989Role-of-renal-i}. Concerning the four case studies of Section~\ref{sec:select}, cases 2 and 3 are in best agreement with the experimental observations in~\cite{Garcia-Estan1989Role-of-renal-i,Khraibi2001Role-of-gender-,Skarlatos1994Spontaneous-cha,Khraibi2000Renal-interstit,Majid2001Nitric-oxide-de,Granger1988Effects-of-rena}. In contrast, case 4 deviates from the experimental observations.

\begin{figure}[tb]
\centering
\begin{center}
        \includegraphics [scale=0.32] {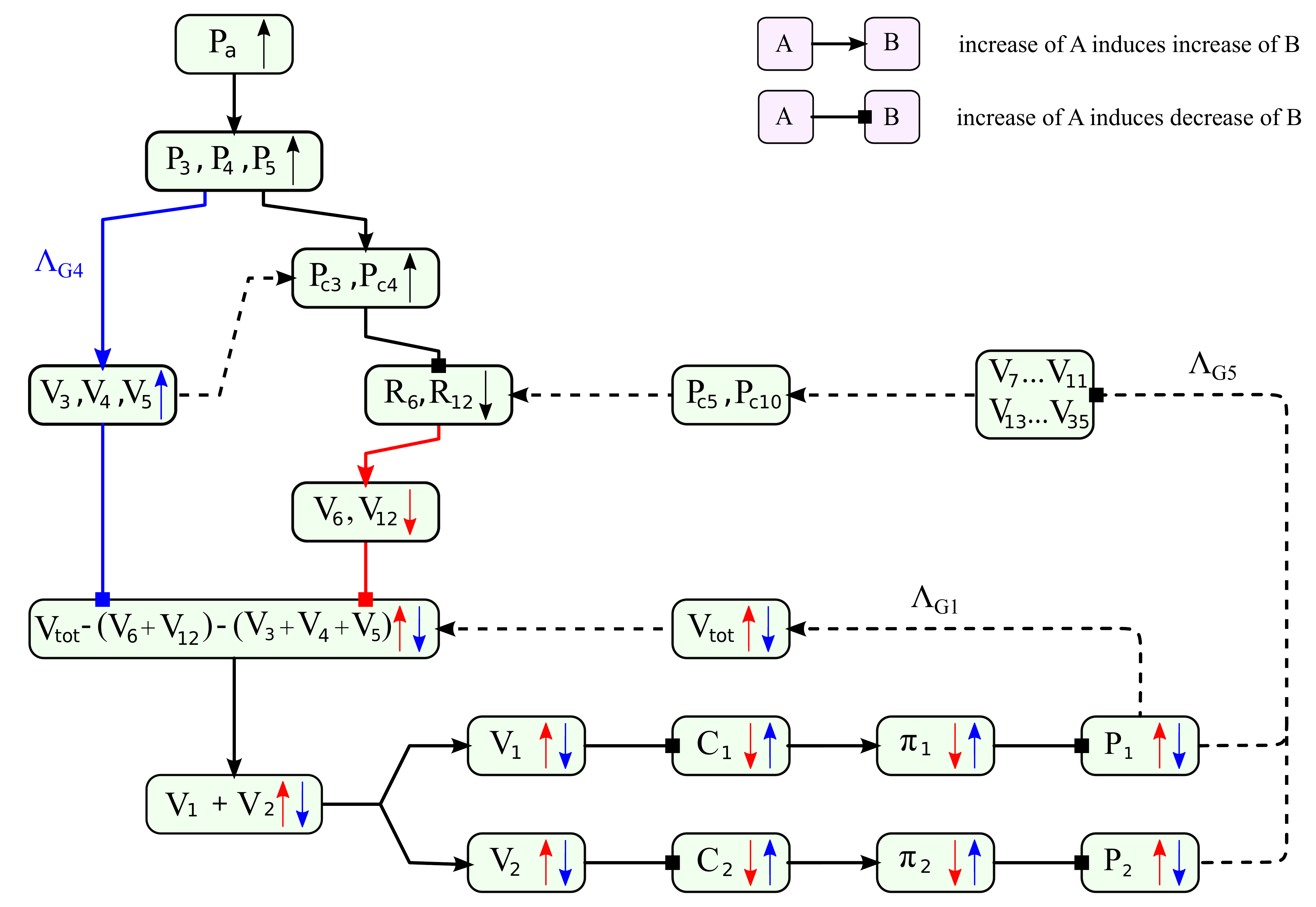}
        \caption{A summary of the mechanism relating arterial blood pressure $P_a$ and interstitial pressures $P_1$ and $P_2$. Changes in $P_a$ are transmitted to $P_1$ and $P_2$ primarily by two pathways: one is mediated by afferent arteriole volumes ($V_6$, $V_{12}$) which is marked with red arrows, the other is mediated by pre-afferent arteriole volumes ($V_3$, $V_4$, $V_5$) and is marked with blue arrows. The two pathways have competing effects. Secondary interactions are denoted with dashed lines. For simplicity, some of the secondary interactions are omitted.}
        \label{fig:block_summary}
\end {center}
\end{figure}

As arterial blood pressure $P_a$ increases, mainly two distinct pathways that lead to interstitial pressure $P_1$ and $P_2$ changes can be identified (Figure~\ref{fig:block_summary}). The first pathway (denoted with red) leads to \emph{increase} of interstitial pressure upon constriction of the afferent arterioles. The second pathway (denoted with blue) leads to \emph{decrease} of interstitial pressure upon dilation of the pre-afferent arteriole blood vessels. Primarily, both pathways lead to changes in interstitial volumes $V_1$ and $V_2$, which are subsequently transmitted to protein concentrations $C_1$ and $C_2$, oncotic pressures $\pi_1$ and $\pi_2$, and finally to $P_1$ and $P_2$.  The two pathways have competing effects; the first leads to changes of $P_1$ and $P_2$ towards the same direction as $P_a$, while the second leads to changes of $P_1$ and $P_2$ towards the opposite direction of $P_a$. It is important to note that, in general, both pathways are active. However, the model results (Figure~\ref{fig:hist_7_factors}) indicate that under most circumstances the first pathway dominates over the second.

The model predictions appear particularly sensitive to the flexibility of the pre-afferent arteriole blood vessels (histological group G4) (Figure~\ref{fig:indices_7_factors}). Such behavior is attributed mostly to the fact that blood pressure is only regulated by the afferent arterioles, which are located after these vessels \cite{Sgouralis2015Mathematical-mo}. The lack of pressure regulation, in the pre-afferent arteriole compartments,  leads to larger internal pressure $P_i^{int}$ changes upon increases in arterial pressure $P_a$ than in the rest of the compartments. For example, as $P_a$ increases from 100~mmHg to 180~mmHg, assuming an increase in the interstitial pressures of $\sim$5~mmHg, we see that the compartments of group G4 are stretched by a pressure difference of $\sim$70--75~mmHg, while the walls of the rest of the compartments are stretched by a pressure difference of $\sim$5~mmHg. Thus, in view of the pressure-volume relations given by equation~\eqref{eq:pressure_volume} the resulting change in total kidney volume $V_0$, which mediates the changes in interstitial pressures, is mostly affected by $s_{G4}$ rather than $s_{G1}$, $s_{G2}$, $s_{G3}$, or $s_{G5}$.

The model developed in this study uses several simplifications. For example, the current model assumes perfect autoregulation of blood flow for equations~\eqref{eq:pois_AA_1}--\eqref{eq:pois_AA_2}, which limits its applicability to cases with arterial blood pressures between 80~mmHg and 180~mmHg \cite{Sgouralis2015Mathematical-mo}. The model does not account for the differences in tubular reabsorption, e.g.\ coefficients $f_i$ in \eqref{eq:flux}, occurring between diuretic and antidiuretic animals or for pressure-diuretic responses \cite{Cowley1997Role-of-the-ren,Moss2013Hormonal-regula}. Further, the model assumes linear pressure-volume relationships for equations~\eqref{eq:pressure_volume} and \eqref{eq:capsule}. Lifting those limitations requires a more detailed model, the development of which will be the focus of future studies. Despite these limitations, the present model could be a useful component in comprehensive models of renal physiology.

%%%%%%%%%%%%%%%%%%%%%%%%%%%%%%%%%%%%%%%%%%%%%%%%%%%%%%%%%%%%%%%%
%%%%%%%%%%%%%%%%%%%%%%%%%%%%%%%%%%%%%%%%%%%%%%%%%%%%%%%%%%%%%%%%
\section*{Acknowledgment}

The authors thank Dr.\ Vasileios Maroulas for assistance with the statistical analysis in this study and for other helpful discussions. This work is conducted as a part of the 2015 Summer Research Experience for undergraduates and teachers at the National Institute for Mathematical and Biological Synthesis (NIMBioS), sponsored by the National Science Foundation through NSF Award $\#$DBI-1300426, with additional support from The University of Tennessee, Knoxville.

% , , , , , . 

\begin{table}[tbp]
\begin{center}
\begin{tabular} { c l c  c  c  c  c  c c}
$i$ 	& Compartment 						& Type 	& Number 	& $P^{int}_i$	& $P^{ext}_i$ 			& Nodes 		& Frac. Coeff.\\
\hline\hline
1 	& Cortical interstitium 					& region 	&1			& $P_1$ 		& - 					& - 			& -  \\
2 	& Medullary interstitium 					& region 	&1			& $P_2$ 		& - 					& - 			& - \\
3 	& Medullary artery 						& pipe 	&8			& $P_3$ 		& $P_2$ 				& c1-c2 		& 0\\
4 	& Arcuate artery 						& pipe 	&24			& $P_{4}$ 	& $\frac{P_1+P_2}{2}$ 	& c2-c3 		& 0\\
5 	& Cortical radial artery 					& pipe 	&864			& $P_5$ 		& $P_1$  				& c3-c4 		& 0\\
6 	& Afferent arteriole\textsuperscript{sn} 		& pipe 	&20736		&  $P_6$ 		& $P_1$ 				& c4-c5 		& 0\\
7 	& Glomerular capillary\textsuperscript{sn} 		& pipe 	&5598720		&  $P_7$ 		& $P_{c19}$			& c5-c6  		&  $\sfrac{3}{28}$ \\
8 	& Efferent arteriole\textsuperscript{sn}  		& pipe 	&20736		&  $P_8$ 		& $P_1$				& c6-c7 		& 0\\
9 	& Cortical capillary 						& pipe 	&1658880		& $P_9$ 		& $P_1$				& c7-c8 		& see~Eq.~\eqref{eq:J9}\\
10	& Venule\textsuperscript{sn}  				& pipe 	&20736		& $P_{10}$ 	& $P_1$				& c8-c9 		& 0\\
11 	& Cortical radial vein 					& pipe 	&864			&  $P_{10}$ 	& $P_1$				& c9-c16		& 0\\
12 	& Afferent arteriole\textsuperscript{ln}  		& pipe 	&10368		& $P_{12}$ 	& $P_1$				& c3-c10		& 0\\
13 	& Glomerular capillary\textsuperscript{ln}  		& pipe 	&4302720		&  $P_{13}$ 	& $P_{c24}$			& c10-c11		& $\sfrac{3}{28}$\\
14 	& Efferent arteriole\textsuperscript{ln}  		& pipe 	&10368		&  $P_{14}$ 	& $P_1$				& c11-c12 	& 0\\
15 	& Descending vas rectum 					& pipe 	&207360		&  $P_{15}$ 	& $P_2$				& c12-c13 	& 0 \\
16 	& Medullary capillary						& pipe 	&10368000	& $P_{16}$ 	& $P_2$ 				& c13-c14 	& see~Eq.~\eqref{eq:J16}\\
17 	& Ascending vas rectum 					& pipe 	&414720		&  $P_{17}$ 	& $P_2$				& c14-c15 	& 0\\
18 	& Venule\textsuperscript{sn} 				& pipe 	&10368		&  $P_{18}$ 	& $P_1$				& c15-c16 	& 0\\
19 	& Arcuate vein 							& pipe 	&24			&  $P_{19}$ 	& $\frac{P_1+P_2}{2}$	& c16-c17 	& 0\\
20 	& Medullary vein 						& pipe	&8			& $P_{20}$ 	& $P_2$				& c17-c18 	& 0\\
21 	& Glomerulus\textsuperscript{sn}  			& sphere	&20736		&  $P_{c19}$  	& $P_1$ 				& c19 		& -\\
22 	& Proximal tubule\textsuperscript{sn}  		& pipe 	&20736		&  $P_{22}$ 	& $P_1$				& c19-c20 	& $\sfrac{2}{3}$\\
23 	& Descending limb\textsuperscript{sn} 		& pipe 	&20736		& $P_{23}$ 	& $P_2$				& c20-c21 	& $\sfrac{3}{10}$\\
24 	& Medullary ascending limb\textsuperscript{sn} 	& pipe 	&20736		& $P_{24}$ 	& $P_2$ 				&  c21-c22 	& 0\\
25 	& Cortical ascending limb\textsuperscript{sn}  	& pipe 	&20736		&  $P_{c24}$ 	& $P_1$ 				& c22-c23 	& 0\\
26 	& Distal tubule\textsuperscript{sn}  			& pipe  	&20736		&  $P_{26}$  	& $P_1$				& c23-c29 	& $\sfrac{13}{84}$ \\
27 	& Glomerulus\textsuperscript{ln}  			& sphere 	&10368		& $P_{24}$ 	& $P_1$ 				& c24 		& -\\
28 	& Proximal tubule\textsuperscript{ln} 			& pipe 	&10368		& $P_{28}$  	& $P_1$ 				& c24-c25 	& $\sfrac{2}{3}$\\
29 	&  Descending limb\textsuperscript{ln} 		& pipe 	&10368		& $P_{29}$ 	& $P_2$ 				& c25-c26 	& $\sfrac{5}{12}$\\
30 	& Medullary ascending limb\textsuperscript{ln}  	& pipe 	&10368		&  $P_{30}$ 	& $P_2$ 				& c26-c27 	& 0\\
31 	& Cortical ascending limb\textsuperscript{ln}  	& pipe 	&10368		& $P_{31}$  	& $P_1$ 				& c27-c28 	& 0\\
32 	&  Distal tubule\textsuperscript{ln} 			& pipe 	&10368		&  $P_{32}$  	& $P_1$ 				& c28-c29 	& 0\\
33 	& Cortical collecting duct 					& pipe 	&144			& $P_{33}$  	& $P_1$				& c29-c30 	& $\sfrac{13}{84}$\\
34 	& Medullary collecting duct 				& pipe 	&144			& $P_{34}$  	& $P_2$ 				& c30-c31 	& $\sfrac{12}{13}$\\
35 	& Papillary collecting duct 					& pipe 	&8			&  $P_{35}$ 	& $P_2$ 				& c31-c32 	& 0\\
\hline

\end{tabular}

\caption{Summary of the compartments contained in the kidney model. Superscripts \emph{sn} and \emph{ln} denote short and long nephrons, respectively. Number refers to the total number of compartments contained in the full model.
}

\label{tb:1}
\end{center}
\end{table}

\begin{table}[tbp]
\begin{center}
\begin{tabular} { cccccccc||cc }
$i$	& $L_i$				&$\mu_i$	& $P_i^{ref}$	&$\Delta P^{ref}_i$	&$R^{ref}_i$		&  $V^{ref}_i$	& $\tilde\sigma_i$		&	$ci$	&	$P_{ci}^{ref}$\\
	&$\mu$m				&		&mmHg		&mmHg			&$\mu$m			&$\mu$m$^3$	& 					&			& mmHg\\
\hline\hline
1	&-					&-		&6			&-			&-		&7.62$\times$10$^{10}$		&-					&	c1		&	100\\
2	&-					&-		&6			&-			&-		&4.92$\times$10$^{10}$		&-					&	c2		&	97.51\\
3	&7$\times$10$^3$		&$\mu_L$	&98.75		&-92.75		&270 	&1.60$\times$10$^9$		&$\tilde\sigma_{G4}$	&	c3		&	95.02\\
4	&2$\times$10$^3$		&$\mu_L$	&96.26		&-90.26		&150 	&1.41$\times$10$^8$		&$\tilde\sigma_{G4}$	&	c4		&	93.97\\
5	&3$\times$10$^3$		&$\mu_L$	&94.50		&-88.50		&75		& 5.30$\times$10$^7$		&$\tilde\sigma_{G4}$	&	c5		&	51.17\\
6	&300	        				&$\mu_A$	&72.57		&-66.57		&10		&9.42$\times$10$^4$		&-					&	c6		&	48.08\\
7	&80	 				&$\mu_C$&49.62		&-37.27		&4.2	 	& 4.43$\times$10$^3$		&$\tilde\sigma_{G5}$	&	c7		&	14.38\\
8	&310	 				&$\mu_E$	&31.23		&25.23		&11	 	&1.17$\times$10$^5$		&$\tilde\sigma_{G5}$	&	c8		&	8.92\\
9	&40	 				&$\mu_C$&11.65		&-5.65		&4.2	 	&2.21$\times$10$^3$		&$\tilde\sigma_{G5}$	&	c9		&	5.44\\
10	&50					&$\mu_L$	&7.17		&-1.18		&12	 	&2.26$\times$10$^4$		&$\tilde\sigma_{G5}$	&	c10		&	50.52\\
11	&3$\times$10$^3$		&$\mu_L$	&5.40		&0.60		&150	 	&2.12$\times$10$^8$		&$\tilde\sigma_{G5}$	&	c11		&	47.51\\
12	&260					&$\mu_A$	&72.77		&-66.77		&10		&8.16$\times$10$^4$		&-					&	c12		&	12.94\\
13	&100	 				&$\mu_C$&49.02		&-35.35		&4.2		&5.54$\times$10$^3$		&$\tilde\sigma_{G5}$	&	c13		&	9.88\\
14	&265	 				&$\mu_E$	&30.22		&-24.22		&11	 	&1.00$\times$10$^5$		&$\tilde\sigma_{G5}$	&	c14		&	9.12\\
15	&210		 			&$\mu_E$	&11.41		&-5.41		&9	 	&5.34$\times$10$^4$		&$\tilde\sigma_{G5}$	&	c15		&	7.78\\
16	&60					&$\mu_C$&9.50		&-3.50		&4.2	 	&3.32$\times$10$^3$		&$\tilde\sigma_{G5}$	&	c16		&	5.37\\
17	&210					&$\mu_A$	&8.45		&-2.45		&9 	 	&5.34$\times$10$^4$		&$\tilde\sigma_{G5}$	&	c17		&	4.41\\
18	&30		 			&$\mu_A$	&6.58		&-0.58		&12 	 	&1.35$\times$10$^4$		&$\tilde\sigma_{G5}$	&	c18		&	4\\
19	&2$\times$10$^3$		&$\mu_L$	&4.89		&1.11		&190	 	&2.26$\times$10$^8$		&$\tilde\sigma_{G5}$	&	c19		&	12.36\\
20	&7$\times$10$^3$		&$\mu_L$	&4.20		&1.79		&425	 	&3.97$\times$10$^9$		&$\tilde\sigma_{G5}$	&	c20		&	11.73\\
21	&-					&-		&12.36		&-6.36		&80 	 	&2.14$\times$10$^6$		&$\tilde\sigma_{G2}$	&	c21		&	11.30\\
22	&14$\times$10$^3$		&$\mu_N$&12.04		&-6.04		&15	 	& 9.89$\times$10$^6$		&$\tilde\sigma_{G3}$	&	c22		&	10.93\\
23	&2$\times$10$^3$		&$\mu_N$&11.51		&-5.51		&8.5 		& 4.53$\times$10$^5$		&$\tilde\sigma_{G3}$	&	c23		&	10.79\\
24	&2$\times$10$^3$		&$\mu_N$&11.12		&5.11		&8.5		& 4.53$\times$10$^5$		&$\tilde\sigma_{G3}$	&	c24		&	13.66\\
25	&3$\times$10$^3$		&$\mu_N$&10.86		&-4.86		&12		&1.35$\times$10$^6$		&$\tilde\sigma_{G3}$	&	c25		&	12.90\\
26	&5$\times$10$^3$		&$\mu_N$&10.73		&-4.73		&13.5	&2.86$\times$10$^6$		&$\tilde\sigma_{G3}$	&	c26		&	11.76\\
27	&-					&-		&13.66		&-7.66		&100		&4.18$\times$10$^6$		&$\tilde\sigma_{G2}$	&	c27		&	10.84\\
28	&14$\times$10$^3$		&$\mu_N$&13.28		&-7.28		&55		&9.89$\times$10$^6$		&$\tilde\sigma_{G3}$	&	c28		&	10.79\\
29	&5$\times$10$^3$	 	&$\mu_N$&12.33		&-6.33		&8.5		&1.13$\times$10$^6$		&$\tilde\sigma_{G3}$	&	c29		&	10.66\\
30	&5$\times$10$^3$		&$\mu_N$&11.30		&-5.30		&8.5		&1.13$\times$10$^6$		&$\tilde\sigma_{G3}$	&	c30		&	6.64\\
31	&1$\times$10$^3$ 		&$\mu_N$&10.82		&-4.82		&12		&4.52$\times$10$^5$		&$\tilde\sigma_{G3}$	&	c31		&	2.00\\
32	&5$\times$10$^3$		&$\mu_N$&10.73		&-4.73		&13.5	&2.86$\times$10$^6$		&$\tilde\sigma_{G3}$	&	c32		&	2\\
33	&1.5$\times$10$^3$		&$\mu_N$&8.65		&-2.65		&16		&1.20$\times$10$^6$		&$\tilde\sigma_{G3}$	&	\\
34	&4.5$\times$10$^3$		&$\mu_N$&4.32		&1.68		&16		&3.61$\times$10$^6$		&$\tilde\sigma_{G3}$	&	\\
35	&2.5$\times$10$^3$		&$\mu_N$&2.00		&4.00		&2.3		&4.15$\times$10$^{10}$		&$\tilde\sigma_{G1}$	&	\\
\hline
\end{tabular}
\caption{Parameter and reference values for the model compartments (indexed by $i$) and nodes (index by $ci$). \emph{Viscosity values:} $\mu_{L}=6.4\times10^{-7}$~min$\cdot$mmHg, $\mu_{A}=2\times10^{-6}$~min$\cdot$mmHg, $\mu_{E}=2.5\times10^{-6}$~min$\cdot$mmHg, $\mu_{C}=4.9\times10^{-6}$~min$\cdot$mmHg, and $\mu_{N}=5.4\times10^{-8}$~min$\cdot$mmHg. \emph{Flexibility values:} $\tilde\sigma_{G1}=0.002$~mmHg$^{-1}$, $\tilde\sigma_{G2}=0.005$~mmHg$^{-1}$, $\tilde\sigma_{G3}=0.045$~mmHg$^{-1}$, $\tilde\sigma_{G4}=0.004$~mmHg$^{-1}$, $\tilde\sigma_{G5}=0.065$~mmHg$^{-1}$.}
\label{tb:3}
\end{center}
\end{table}

\end{document}